\documentclass[aps,pre,twocolumn,amssymb,showpacs,groupedaddress]{revtex4-2}

\usepackage{graphicx,color}
\usepackage[normalem]{ulem}
\usepackage{hyperref}
\usepackage{amsmath}
\usepackage{tabularx}
\usepackage{comment}
\usepackage{appendix}
\def\beq{\begin{equation}}
\def\eeq{\end{equation}}
\def\bea{\begin{eqnarray}}
\def\eea{\end{eqnarray}}

\begin{document}
\definecolor{colortodo}{RGB}{255,0,0}
\newcommand{\red}[1]{{\color{colortodo}#1}}
% number in the upper righthand corner of the title page in preprint 
%\preprint{}
\title{Memory in cyclically crumpled sheets}
% Please use the appropriate macro for each type of information
\author{Amit Dawadi}
\affiliation{Department of Physics, Clark University, Worcester, Massachusetts 01610, USA}
\author{Arshad Kudrolli}
\affiliation{Department of Physics, Clark University, Worcester, Massachusetts 01610, USA}
\date{\today}
\begin{abstract}
We investigate the crumpling of an elastoplastic sheet as it is repeatedly crushed onto itself by rolling it into a cylinder and twisting it axially while allowing the end-to-end length to evolve freely. The sheet buckles and folds into structures which repeat but sharpen over hundreds of cycles to a remarkable degree before forming different configurations. Below a critical amplitude, reconfigurations decrease with applied cycles but continue to occur for large enough loading amplitude as the topology of the sheet changes as it tears. The sheet structure as measured by the mean curvature and the total crease length evolves logarithmically with cycle number with a rate which increases with compaction. We explain the progress of creasing using a flat folding model, and show the logarithmic growth as being a consequence of individual creases becoming sharper with number of folding cycles, leading to bifurcations in the folding pathway. 
\end{abstract}
\maketitle

\section{Introduction}
A crumpled sheet shows a distinct pattern of intersecting creases which carry the subtle imprints of the complex buckling pathways followed as a result of applied constraints~\cite{Amar1997,Blair2005,Boue2006,Witten2007,Croll2019}. The evolution of the observed disordered ridges and facets can provide broad lessons on a system that can typically only access metastable states~\cite{Matan2002,Fokker2019,Venkataramani2019}. While there have been considerable number of experimental and theoretical studies on memory effects after the application or removal of stress~\cite{Persson1998,Bertin2003,Thiria2011,Amir2012}, the evolution under repeated loading cycles has only started garnering attention more recently. As a result of sensitive dependence on the loading conditions, a sheet cannot be expected to pass the same folding path if it is crumpled again, thus leading to further creases and hysteresis~\cite{Gottesman2018,Keim2019}. Nonetheless, the presence of creases influences how the sheet subsequently collapses when the loading cycle is repeated as the sheet rigidifies along the length of the crease, while becoming easier to fold about the crease~\cite{DiDonna2001}, leading to guided folding pathways as in origami~\cite{Santangelo2017}, and possibly their Poynting response~\cite{Ribes2024}. Indeed, supervised strengthening folds and weakening of misfolds in a repeatedly crumpled sheet has been suggested as a paradigm for reinforced learning of structures in mechanical systems subjected to classes of forces~\cite{Stern2020}.  It has been argued based on numerical simulations, that the energy landscape of a folded sheet can become deeper under repeated loading, driving the sheet to unique folding pathways
~\cite{Stern2018}. However, investigations with actual sheets remain limited~\cite{Arinze2023}. 

In athermal granular systems, reversible motion and relaxation to limit cycle can be observed in cyclically sheared colloidal suspensions and amorphous solids that show irreversible motion above a critical amount of deformation~\cite{Pine2005,Regev2013}. Hysteresis and an approach to a limit cycle have been observed in pre-crumpled sheets when repeated loading cycles are below the threshold required to create additional crumples~\cite{Shohat_2022}. But, even a pre-creased ordered origami can snap between different deployed structures because of their multistability~\cite{Jules2022,Xia2022,Dunne2022}. Furthermore,  
folds in elastoplastic sheets can age~\cite{Thiria2011,Jules2022} impacting their compaction properties~\cite{Tallinen2009,Habibi2017}. Thus, the evolution of the crumpled structure of a sheet subject to large number of repeated loading and unloading cycles of varying strength, remains unclear.

To address the evolution of an elastoplastic sheet subject to large number of training cycles, we examine the crumpling of a Mylar sheet which is clamped onto circular end caps, and then repeatedly crushed by applying an axial twist akin to wringing a towel~\cite{Rahmayanti2016}. Because of the inextensible nature of the sheet, its length contracts axially, and the sheet collapses onto itself while forming a bundle~\cite{Dong2023}  in the strongly crumpled regime where the sheet comes in self-contact~\cite{Houle1996}. A hallmark of our design is that the applied strain can be reversed back to zero, and thus the loading cycle can be repeated essentially indefinitely under well prescribed conditions. This system enables us to examine the effect of sheet training over a wide range of compaction before material fatigue leads to fractures or tears in the sheet. We highlight the strong effect that repeated folding and unfolding elastoplastic sheets has on crease evolution relative to ageing.  

\section{Experimental methods}
\label{sec:setup}
%\subsection{Materials and apparatus}
\begin{figure}[b]
    \centering
    \includegraphics[width=7cm]{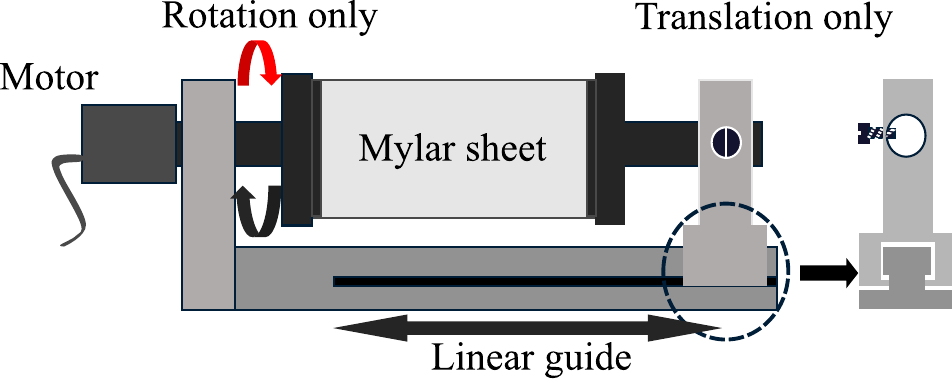}
    \caption{ The apparatus used to apply cyclic twist to the sheet. The dashed circle indicates the ball bearing sliders which allow essentially frictionless translation and prevent rotation.}
    \label{fig:setup}
\end{figure}
We study sheets composed of a biaxially oriented polyethylene terephthalate (BoPET), also known as Mylar, of length $L_0=16.5$\,cm, width $W_0 =16.5$\,cm, and thickness $h_0 = 90$\,$\mu$m. These sheets have a bending modulus $E = 4.16$\,GPa, obtained with standard beam bending measurements. The sheet is rolled and clamped onto two circular rigid aluminum disks of diameter $D=52.5$\,mm to form a cylinder of length $L_0$  as shown in Fig.~\ref{fig:setup}. Then one end is twisted axially by a computer controlled Parker stepper motor through an angle $\theta$ with a rate $\omega = 3.67$\,degrees/s (see Supplementary Movie~S1~\cite{sup-doc}). Thus, a loading cycle takes approximately 6~minutes and 30~seconds when $\theta$ is increased to $\theta_f = 720^\circ$ and then returned back to $0^\circ$. The other end is not allowed to twist, but allowed to move along its axis while mounted on linear guides with coefficient of friction less than $0.1$. Although an axial load can be prescribed, we perform measurements while allowing the sheet to contract or expand freely.

%\subsection{Laser profilometry}
The surface profile of the sheet is obtained by illuminating the sheet with a 640\,nm red laser sheet orthogonal to the unperturbed surface. The illuminated surface is then imaged at an angle with a Pixelink digital camera with a resolution of $2592 \times 2048$\,pixels. The bright pixels are used to locate the surface with a centroid algorithm to within a few microns. This method yields a single height profile along the length of the sheet while mounted on the apparatus. To obtain a full scan of the sheet surface, the sheet is unmounted to relieve elastic stresses. It is then placed on a flat scanner bed after being flattened somewhat to remove overhangs. Care is taken to minimize introducing plastic deformations in the process. Because the sheet cannot be remounted to have the same boundary conditions as when it was dismounted, we start with a fresh sheet for each scan while measuring as a function of cycle number $n$ or final twist angle $\theta_f$. 

\begin{figure*}
\centering
\includegraphics[width=18cm]{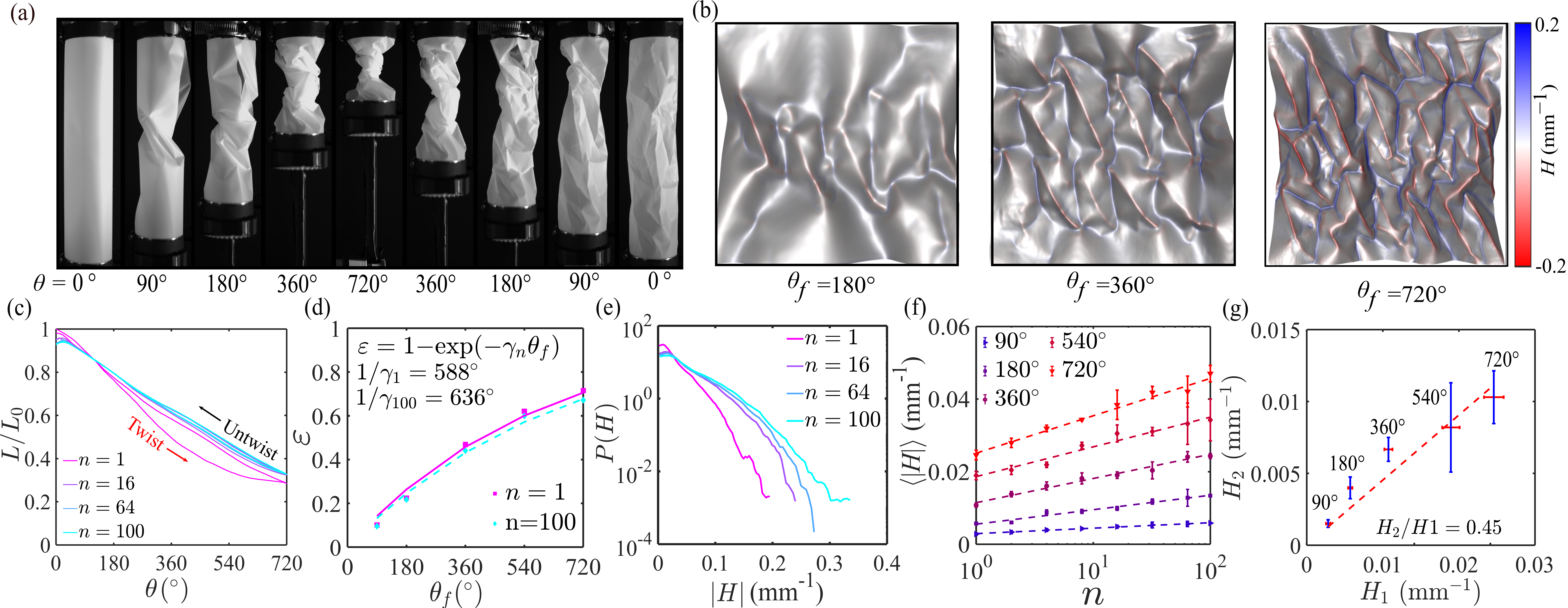}
\caption{%Color online. 
(a) Images of a Mylar sheet as it is twisted and then untwisted about its axis
($L_0=16.5$\,cm%\sout{, $W_0 =16.5$\,cm, and $h= 90$\,$\mu$m}
). (b) The crumples observed after one cycle corresponding to various $\theta_{f}$. The mean curvature $H$ are superimposed with values denoted by the color bar.  (c) The end-to-end length $L$ as a function of $\theta$ shows hysteresis. The area enclosed decreases with $n$. (d) The strain $\varepsilon$ versus $\theta_f$ compared with the 
function $\varepsilon = 1 - \exp(-\gamma_n \theta_f)$. (e) The distribution of $|H|$ broadens with $n$. (f) The evolution of $\langle |H| \rangle$ with $n$ is described by a logarithmic function $\langle |H| \rangle = H_1 + H_2\log n$. (g) $H_1$ and $H_2$ increase with $\theta_f$.} 
\label{fig:twist}
\end{figure*}

\section{Crumpling by twisting}
Figure~\ref{fig:twist}(a) shows the evolution of a sheet 
as it is twisted and then untwisted slowly through various twist angles $\theta$. 
The sheet is observed to buckle with folds slanted along the direction of twist in the central section between the clamps ends, while the sections close to the clamps remain relatively cylindrical~\cite{Donnell1935,HUNT2005,Wang2020} (also see supplementary documentation~\cite{sup-doc}.) As the twist is increased further,  the initial folds break into smaller sections, self-contact occurs, and the sheet crumples and collapses inward along its entire length. When the applied twist is reversed, the crumpled structure show auxetic behavior~\cite{Ribes2023} with increases in both length and diameter as the sheet returns back close to its initial cylindrical shape as the elastic components of folds relax while showing imprints of plastic deformation. 

Figure~\ref{fig:twist}(b) shows scans of the sheet for increasing final twist angle $\theta_f$ using laser profilometry as discussed further in %the Supplementary Document~\cite{sup-doc}
Appendix~\ref{sec:setup}. The local mean curvature of the surface $H$ is also superposed in colors.  In the case of $\theta_f = 180^\circ$, imprints of the initial folds that form can be observed even as the sheet comes into self-contact and the folds start to bend and break to accommodate the increasing twist. The sheet is observed to become more widely creased and disordered as  $\theta_f$ is increased to $360^\circ$ and $\theta_f = 720^\circ$, unlike hyperelastic sheets which show ordered folds~\cite{Chopin2022}. As quantified further in Appendix~\ref{sec:curv}, the mean curvature $\langle|H|\rangle$ near the clamped region is systematically lower compared to the central regions which form a tight bundle. This effect can be seen most clearly in the case of $\theta_f = 180^0$, but is also present at higher $\theta_f$.   

We plot the end-to-end length $L$ of the sheet from one clamp edge to the other in Fig.~\ref{fig:twist}(c)  as a function of $\theta$, and observe that it follows a different path while untwisting, which is systematically higher compared to the twisting phase, before crossing and reaching a length $L_n$ with cycle number $n$ which is slightly below $L_0$. When the sheet is again subjected to the same loading cycle, the hysteresis is observed to decrease on average with $n$. We characterize the compression of the sheet by measuring the end to end length $L_c$ when $\theta_f$ is reached, and plotting the axial strain $\varepsilon = 1 - L_{c}/L_0$ averaged over $3$ trails versus $\theta_f$ in Fig.~\ref{fig:twist}(d). We observe that its increase can be described with the function $\varepsilon = 1 - \exp(- \gamma_n \theta_f)$, where $\gamma_n$ is a fitting constant. This fit is consistent with a sheet becoming increasingly more difficult to compress with increasing compaction, and increasing cumulative plastic deformations with applied cycles, and corresponds to a regime where $L/h_0 \gg 1$.
Comparing images of the crumpled structure after application of a full cycle shown in Supplementary Movie~\cite{sup-doc}, we observe that the creased structures largely repeat with $n$, but slow evolution of overall structure, switching between bi-stable regions, and formation of fresh creases can be also observed. To quantify the evolution with $n$, we obtain the distributions of the magnitude of the local mean curvature $P(|H|)$ and plot them in Fig.~\ref{fig:twist}(e) for various $n$. We observe that it is broadly distributed, and broadens further as the sheet is repeatedly twisted showing that the sheet continues to evolve with $n$ even as $L$ versus $\theta$ graph appears to show decreasing hysteresis (see supplementary documentation~\cite{sup-doc}).

We plot the average of the distribution $\langle |H| \rangle$ of the flattened scanned sheet in Fig.~\ref{fig:twist}(f) and observe that $\langle |H| \rangle$ grows systematically higher with increasing $\theta_f$. Further, each $\langle |H| \rangle$ is well described by a logarithmic function $\langle |H| \rangle = H_1 + H_2 \log n$, where $H_1$ is the magnitude of the mean curvature observed after the application of the first cycle, and $H_2$ captures the growth with number of cycles. As shown in Fig.~\ref{fig:twist}(g), $H_2$ increasing approximately proportional to $H_1$. Thus, we find that not only do the sheet deformations depend on $\theta_f$, but that the rate of their grow with $n$ is essentially proportional the deformations after $n=1$.  The observed trends are relatively insensitive to increasing the loading rates by an order of magnitude (see %supplementary documentation~\cite{sup-doc}
Fig.~\ref{fig:ratedep}).

\section{Logarithmic growth of curvature under repeated loading and loading time}
\begin{figure}
\includegraphics[width=8.5cm]{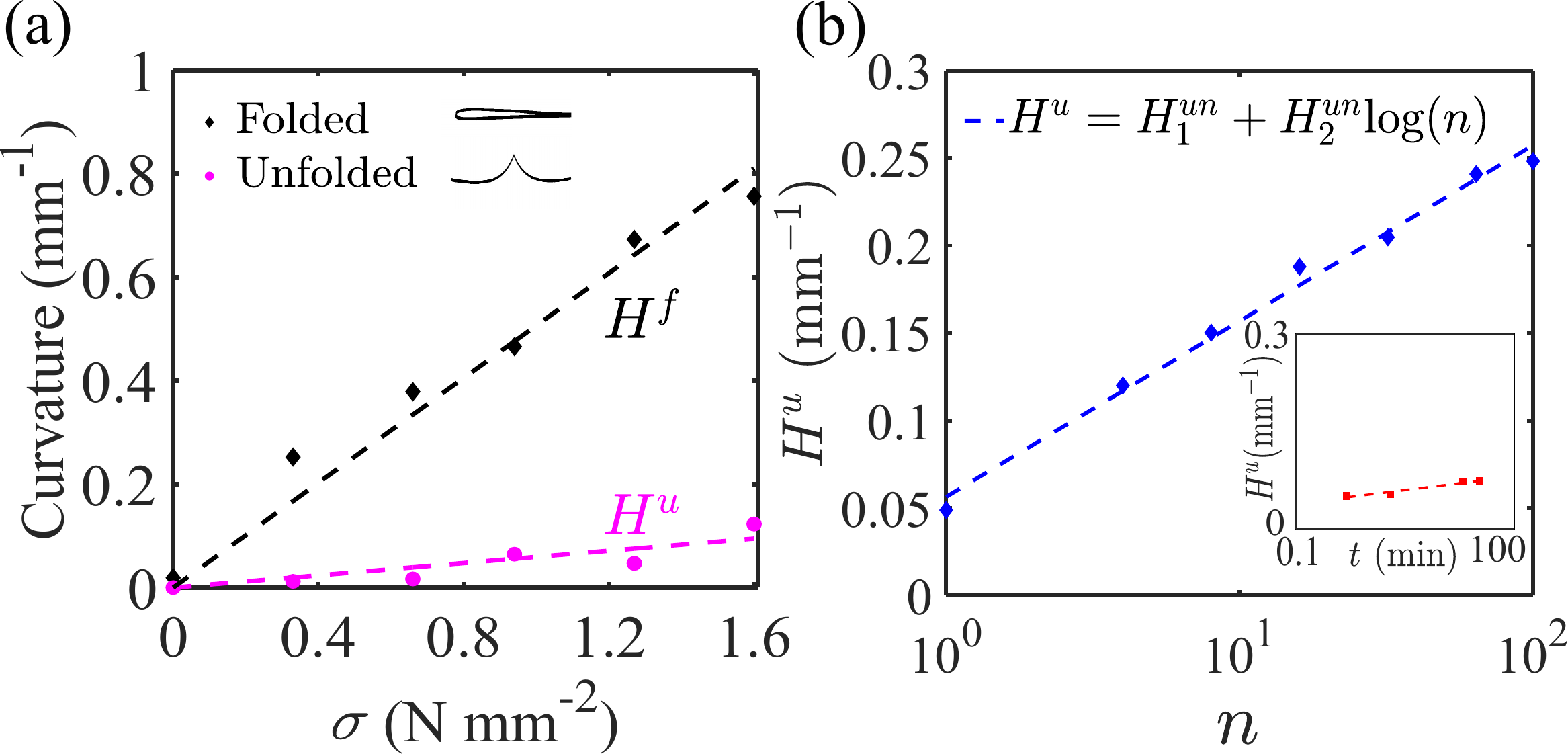}
\caption{%Color online. 
(a) The mean curvature of a folded and a unfolded sheet increase approximately linearly with applied stress. A fit to $H^f = \chi^f \sigma$ and $H^u = \chi^u \sigma$ with $\chi^f=0.5$\,mm\,N$^{-1}$ and $\chi^u=0.06$\,mm\,N$^{-1}$, respectively, are shown. Inset: A cross section of the folded and unfolded sheets are also shown. 
(b) $H^u$ as a function of $n$, and a fit with $H^{un}_1 =0.049$\,mm$^{-1}$ and $H^{un}_2 =0.10$\,mm$^{-1}$  ($\sigma=0.7$\,N\,mm$^{-2}$; $t=30$\,seconds). Inset: $H^u$ increases as a function of time interval over which a constant force is applied. A function $H^u = H^{ut}_1  + H^{ut}_2 \log(t)$ with $H^{ut}_1 =0.051$\,mm$^{-1}$ at t = 30\,seconds and $H^{ut}_2 =0.014$\,mm$^{-1}$min$^{-1}$ is also shown. The variation of $H^u$ is significantly lower over $t$ compared with $n$. 
} 
\label{fig:1ridge}
\end{figure}

We examine the curvature of a single crease upon repeated folding as a starting point to understand the observed logarithmic growth, and any relation to Arrhenius behavior that has been observed in response to application of stress in crumpled sheets~\cite{Thiria2011,Amir2012,Jules2020,Shohat_2023}. Accordingly, a folded sheet is placed between parallel plates and a normal force is applied for a fixed time. The force is then removed  to relax the elastic component of the crease and the unfolded sheet is placed on a flat-bed for scanning. A representative section of the sheet while folded and unfolded is shown in the Inset to Fig.~\ref{fig:1ridge}(a).

We obtain the mean curvature $H^f$ and $H^u$ along the crease while loaded and unloaded, respectively, and plot them in Fig.~\ref{fig:1ridge}(a) as a function of the stress $\sigma$ corresponding to the applied force per unit length of crease and the sheet thickness. We observe that the curvatures for the folded and unfolded sheet evolve linearly as a function of applied stress over the typical range of curvatures observed in our crumpling experiments, similar to observations with copy paper~\cite{Blair2005}. Thus, even after loading is removed, a fraction proportional to the maximum loading applied can be identified, giving us confidence that the physical features of a crushed sheet can be still identified after the sheet is relaxed due to the residual plastic deformations. Next, we perform measurements of 
$H^u$ for a typical crease as a function of number of folding and unfolding cycles for $\sigma = 0.7$\,Nmm$^{-2}$ applied for $30$\,seconds. As shown in Fig.~\ref{fig:1ridge}(b), we find that $H^u$ increases with the number of applied loading cycles $n$, as $H^u = H^{un}_1  + H^{un}_2 \log(n)$ with a rate $ H_2^{un} = 0.10$\,mm$^{-1}$. We also measured the evolution of curvature when the load is applied over 
varying time interval $t$. As shown in Inset to Fig.~\ref{fig:1ridge}(b), we find that the change in $H^u$ with $t$ is significantly lower compared with the increase in $H^u$ with $n$ over comparable cumulative loading time. 

Consequently we find that the repeated folding and unfolding of the crease leads to far greater plastic deformations and is the dominant reason for the logarithmic increase in our experiments. Thus, the evolution in our experiments appears to be more in line with memory observed due to cyclic driving in sheared suspensions~\cite{Pine2005,Paulsen2014}, and  
different from the logarithmic time dependence reported previously in elastoplastic sheets~\cite{Thiria2011,Amir2012,Shohat_2023}.

\section{Evolution of creases under repeated loading}
\begin{figure}
    \centering    
    \includegraphics[width=8cm]{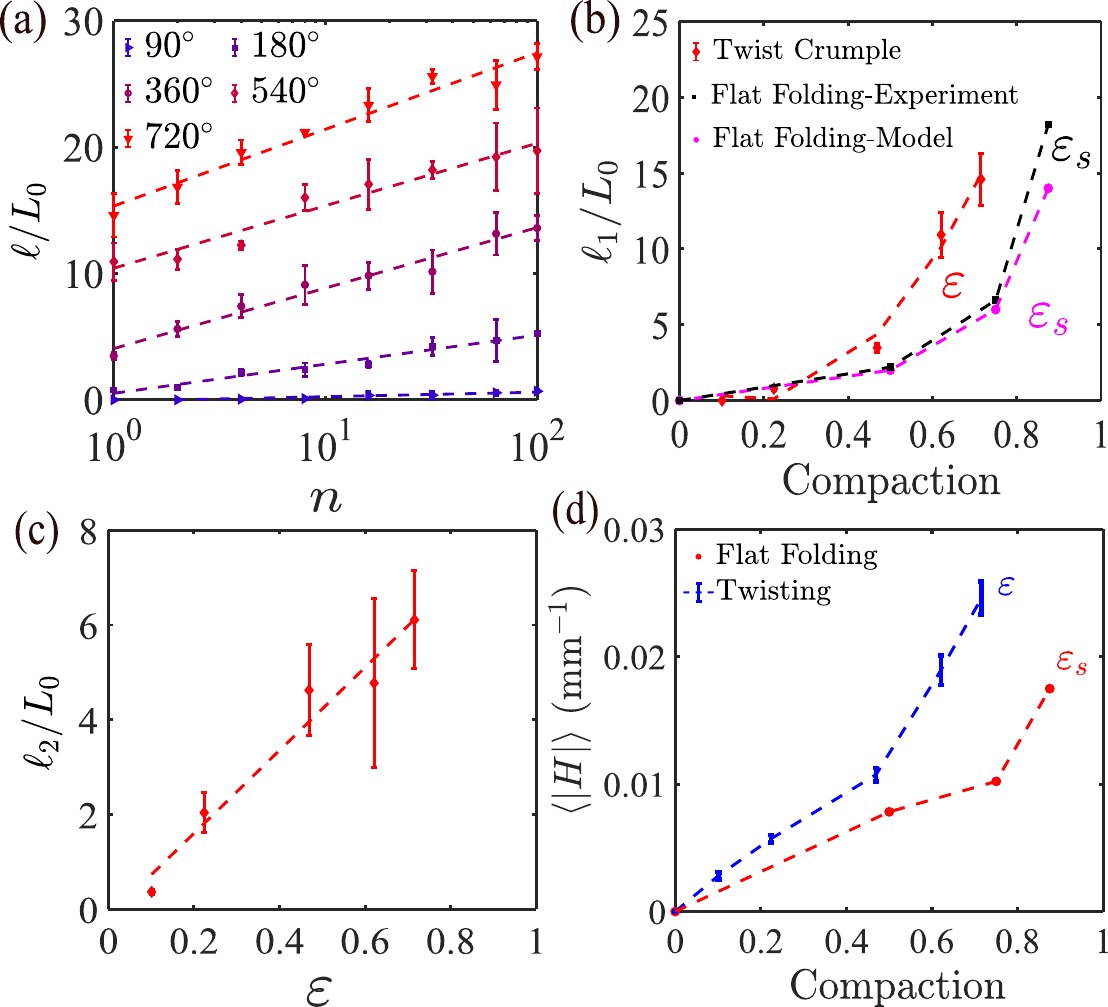}
    \caption{%Color online. 
    (a) The evolution of $\ell$ with $n$ is %\sout{well} 
    described by the logarithmic function $\ell = \ell_1 + \ell_2 \log(n)$. (b) The crease length after the first loading cycle $\ell_1$ increases with compaction $\varepsilon$. The calculated and observed crease lengths when a sheet is folded into increasingly small squares plotted versus compaction $\varepsilon_s$ estimated using the size of the folded sides. (c) The rate of increase of crease length $\ell_2$ also increases with compression. (d)
    The absolute mean curvature of the sheet increases with the compaction in both flat folding and twist crumpling cases.
}
    \label{fig:ridge_length}
\end{figure}

Figure~\ref{fig:twist}(b) shows the regions with high $H$ are organized along intersecting networks of creases with twist crumpling, visually similar to those which have been noted with other crumpling protocols~\cite{Blair2005,Gottesman2018,Cambou2011},  but somewhat anisotropic due to the creases that form following the primary buckling instability (see Appendix~\ref{sec:angle}). We analyze the creases in terms of the total crease length which has been introduced as a measure of crumpling~\cite{Gottesman2018}. The total crease length $\ell$ normalized by $L_0$ as a function of $n$ is plotted in Fig.~\ref{fig:ridge_length}(a) for various $\theta_f$. We observe that it increases in all cases with number of cycles, and can be fitted by logarithm function  $\ell = \ell_1 + \ell_2 \log(n)$, where $\ell_1$ is the total crease length after the first loading cycle and $\ell_2$ is the rate of growth corresponding to the creases with $n$. Both, $\ell_1$ and $\ell_2$ increase with $\varepsilon$, as shown in Fig.~\ref{fig:ridge_length}(b,c), respectively. 

A logarithmic growth of crease length has been reported in sheets crushed uniaxially inside a piston by Gottesman, {\it et al.}~\cite{Gottesman2018}, but with a different two-parameter functional form. In those studies, the sheets were flattened after each cycle to scan the surfaces, and thus random perturbations were introduced which leads to the creation of fresh creases since the start point for the folding sequence is different. Indeed, if we introduce similar perturbations in our system, by unmounting and mounting after flipping the sheet inside out after each cycle, we find logarithmic growth as well, but with far greater rate of increase corresponding to the creation of larger numbers of fresh creases. 

Thus, our study finds that creases grow under repeated loading even when such random perturbations are absent due to the underlying plastic evolution of the sheet curvature under repeated loading. This evolution leads to sharpening curvature of the folds, which then can exceed the thresholds used to identify creases besides formation of new creases when the sheet follows different folding pathways.  

\section{Effect of loading rate and plasticity}
\begin{figure*}
    \centering
    \includegraphics[width=11cm]{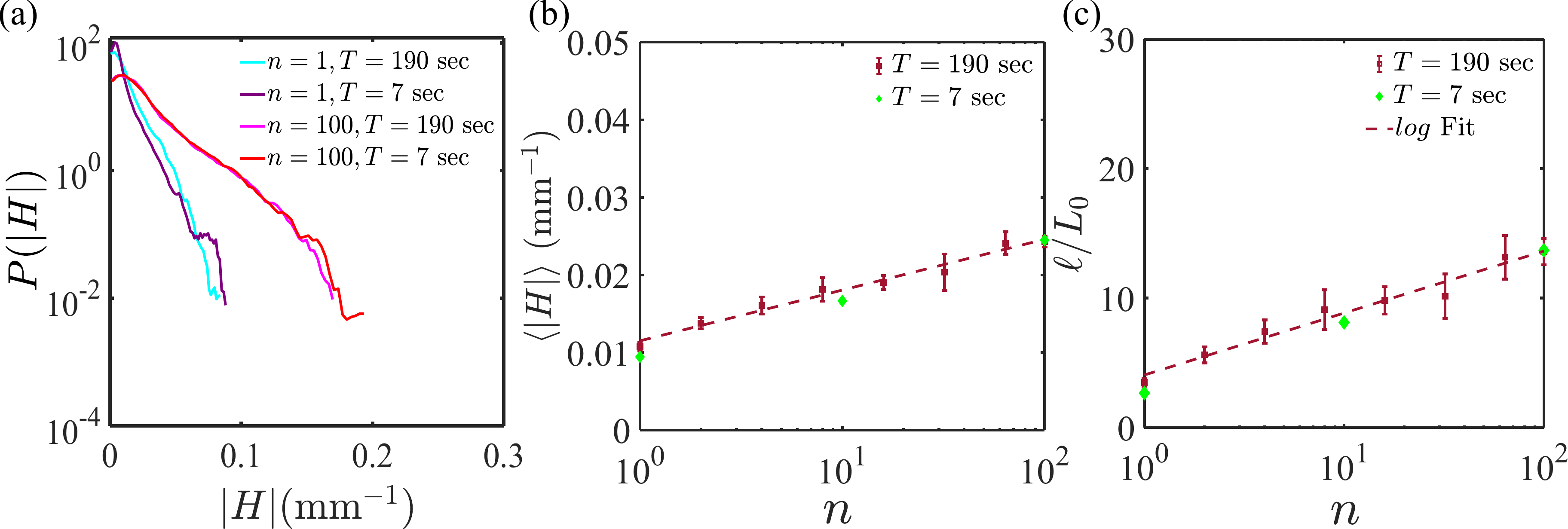}
    \caption{The distribution of $|H|$ (a), $\langle |H| \rangle$ (b), and $\ell$ (c) are  observed to be essentially similar for $360^\circ$ angle of twist as the loading rate is increased from 7 seconds per cycle to 190 seconds per cycle. }
    \label{fig:ratedep}
\end{figure*}

\begin{figure*}
    \centering
    \includegraphics[width=11cm]{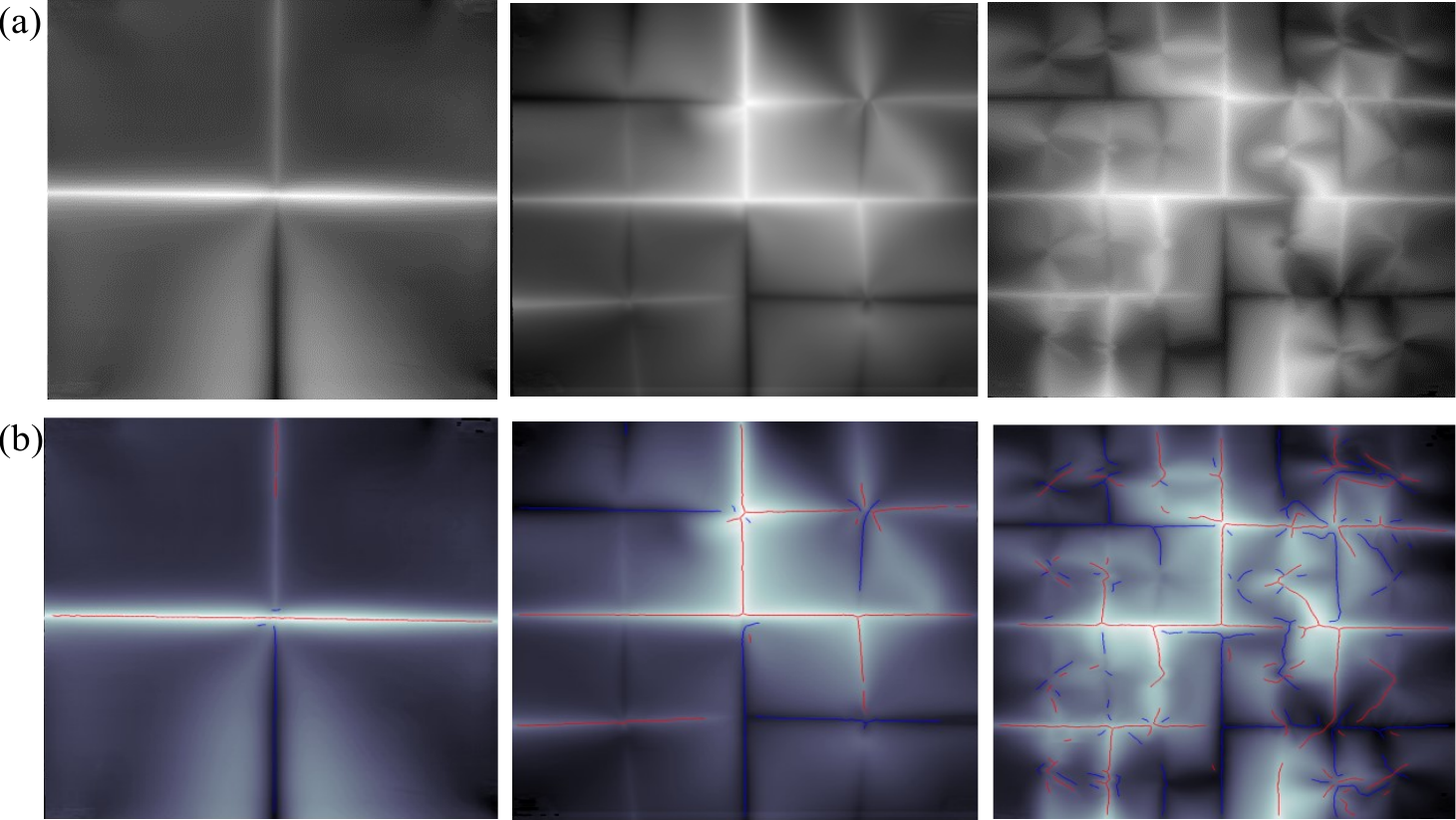}
    \caption{(a) Folds become increasingly disordered as the sheet is folded into smaller squares. (b) The detected ridges and valleys are superimposed on the scans. Each example shown corresponds to starting with a defect free square sheet and applying the folding sequence and then unfolding and scanning.} 
    \label{fig:square}
\end{figure*}

We decreased the loading rate from 190 seconds per cycle to 7 seconds per cycle to study the sensitivity of our measurements to loading rate since plasticity effects can be rate dependent. The distribution of the magnitude of the mean curvature $|H|$ is shown in Fig.~\ref{fig:ratedep}(a) for representative cases of $n=1$ and $n=100$. In both cases, the distributions can be observed to be similar within statistical fluctuations. We plot the mean of the distribution $|H|$ and $\ell$ for the two different rates in Fig.~\ref{fig:ratedep}(b) and Fig.~\ref{fig:ratedep}(c), respectively. The trends in terms of these plots essentially also overlap, showing that the trends shown are relatively insensitive to loading rates probed here. It is not possible to increase the rates much further without adding inertial effects. Currently the experiments take hours to apply hundreds of loading cycles. Decreasing the loading rate by an order of magnitude further in the other direction would imply that each data point would take days, which is practically difficult.

\section{Folding model of creases}
\label{sec:square}

In order to gain an understanding of the overall growth of the creases with applied strain, we study complementary flat square-folding of a sheet into a smaller and smaller volume~\cite{Deboeuf2013}. The scans of folded, unfolded and flattened sheets with the same Mylar sheets as in the twist-crumple experiments are shown in Fig.~\ref{fig:square}(a). The sheet become increasingly more difficult to fold with increasing number of folds, and we found it practically difficult to fold them into still smaller squares. Even though they are creased by an ordered sequence of folds, they show growing disorder and nonuniformity of the curvature along creases.  
Using the same threshold criteria to detect the ridges and valleys as in the twist experiments, we identify them, and superpose them on the scans in Fig.~\ref{fig:square}(b). Examining the detected ridge and valleys, we observe that primary folds, which were missed initially, are detected in their entirety with subsequent folding. However, the new folds corresponding to folding twice and thrice have overall lower curvature because of the thickness of the sheet and are detected to a lower degree.  One also observes the appearance of curvature with opposite signs adjacent to the square folds as the sheet is forced back to a planar configuration. While overall smaller in magnitude, they can be noted to be similar in magnitude to the creases created further along the folding sequence. Further, disorder and creases at the intersection of the creases created while folding, earlier in the sequence, are also observed to appear.  

Complementarily, we also consider a square infinitesimally thin sheet with side $L_0$, which is the same as the sheets used in the twist crumple study. Then, we calculate the total crease length upon flat folding the sheet into smaller squares with sides $L_0/m$, with $m=1,2,4, ..$. This flat folded sheet can fit inside a square area that has a size $L_0/m$. Accordingly, we obtain a compression $\varepsilon_s = 1-1/m$ for flat folded sheets. 

We plot the calculated total crease length of folded sheets and those from the flat-folding experiments in Fig.~\ref{fig:ridge_length}(b), and find them to vary roughly over the same range as the twist crumple experiments, in spite of the differences in the exact definitions of the compaction. Comparing the measured values from flat folding experiments with calculated values assuming ideal folds along lines, we observe that the total length is greater in the case of the experiments. Growing disorder due to the thickness of the sheet results in the formation of secondary creases as can be seen in the scans shown in Fig.~\ref{fig:square}(b). To some extent these creases offset the fact that not all creases are identified, since their unfolded curvatures fall below the threshold used to identify creases. While these two effects offset each other, a somewhat faster growth is observed in the total measured crease length compared to the calculated ideal square folds. 

Figure~\ref{fig:ridge_length}(d) shows the average mean curvature obtained from the flat folding experiments and compare with those obtained by twisting. We find that both increase roughly over the same range, but are systematically higher due to greater disorder in the case of twisting. Thus, this prescribed folding method allows us to not only estimate the total crease length but also gauge practical differences due to folding a sheet with finite thickness.
\section{Metastability}

\begin{figure}
\includegraphics[width=8cm]{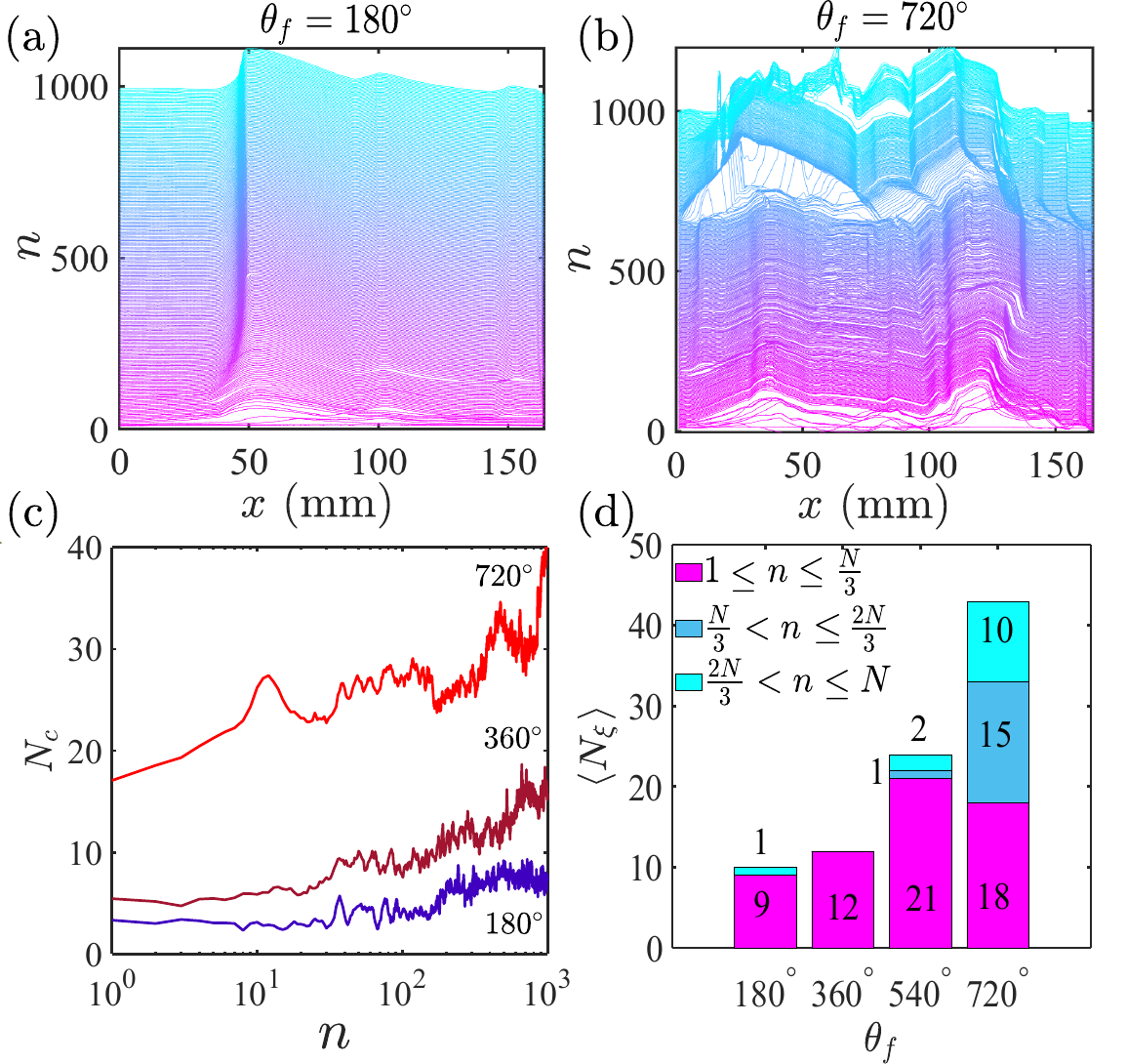}
\caption{%Color online. 
(a,b) The evolution of a sheet cross section shows creases becoming sharper with $n$, sudden changes in the surface profile, and the formation of increasing number of fresh creases  with increasing $\theta_{f}$. (c) The total number of creases $N_c$ along the length of the cylinder counted using the cross-sections measured after each cycle. (d) The number of peaks corresponding to reconfigurations increases with $\theta_f$, but decreases with $n$ for $\theta_f < 720^\circ$.} 
\label{fig:sec_evo}
\end{figure}

Since the full surface scans rely on dismounting the sheet after starting with a pristine sheet, it is only possible to understand statistical trends with those measurements.  As a way to interrelate the crumpled configuration of the sheet after application of a training cycle, we examine the height profile $h(x,n)$ of the sheets along a length-wise section $x$ after each cycle $n$. Representative evolution of a section of the sheet corresponding to $\theta = 180^\circ$  and $720^\circ$ are shown in Fig.~\ref{fig:sec_evo}(a,b), respectively. While the sections evolve rapidly in each case over the first tens of cycles, they converge to show a set of peaks and valleys which recur in the case of $\theta_f = 180^\circ$.  Slow evolution of creases and appearance of fresh creases can be noted  due to the evolving constraints~\cite{Boudaoud2000} even as the sheet peaks and valleys mostly repeat and sharpen with $n$. While, one notes similar trends in the case of $\theta_f = 720^\circ$, sudden rearrangements occur as well, whereby the entire cross-section undergoes a rapid snap-through event. (It must be noted that fresh creases do not have to arise at that location, but changes in the structure elsewhere can also lead the sheet to move relative to the fixed section being examined.)   Because the creases become sharper with increasing number of cycles, the number of creases $N_c$ in $h(x,n)$, identified by using a fixed curvature threshold criterion, continues to grow on average over $n=1000$ even when reconfigurations are absent (see Fig.~\ref{fig:sec_evo}(c)).  

\begin{figure}[h]
    \centering
    \includegraphics[width=7.5cm]{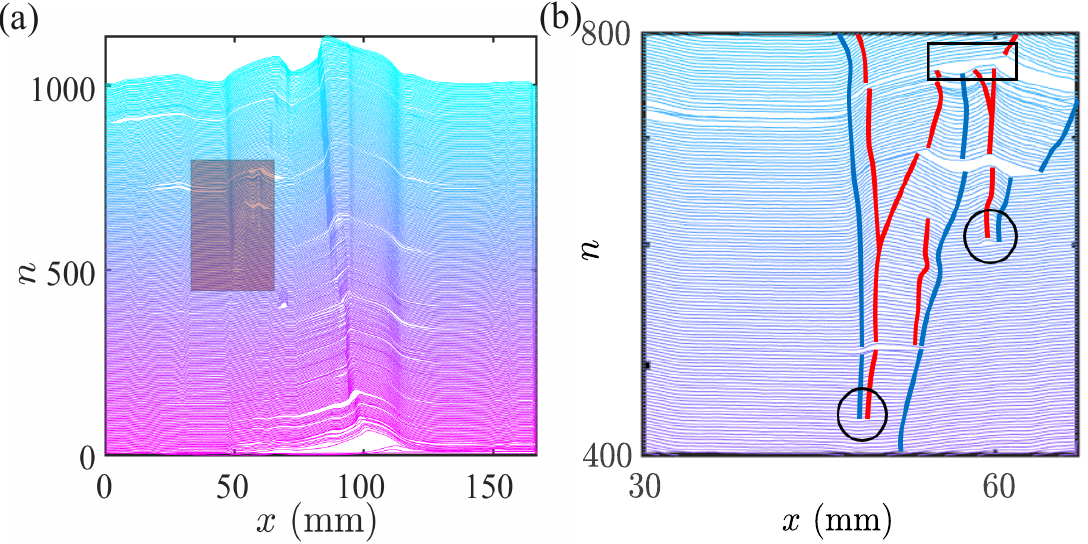}
    \caption{(a) The evolution of a cross-section with $n$ for $\theta_f=360^\circ$. (b) A zoomed in view of the shaded region in Fig.~\ref{fig:ridge_evolution_360}(a) showing the relative motion of the creases with increasing $n$, including creation of peak and valley pairs (black circles) and destruction of creases (black rectangle).}
    \label{fig:ridge_evolution_360}
\end{figure}

These observations further corroborate the discussion of the evolution of a single crease following Fig.~\ref{fig:1ridge}(b), i.e. plastic damage accumulates in regions which have sustained damage in previous cycles. As the overall constraints change, peaks and valleys can be added or removed in the curvature field leading to creation or destruction of creases (see 
Fig.~\ref{fig:ridge_evolution_360}).  Thus, while continuous evolution of the sheet curvature can lead to recurrent folded structures, it can also drive rapid reconfigurations even after the sheet structure repeats over hundreds of cycle as different nearby folding pathways become more favorable. Thus, the picture which emerges from these observations is more subtle than the continuous approach to a limit cycle over tens of cycles shown by Shohat, {\it et al.}~\cite{Shohat_2022} with a precrumpled sheet subject to perturbative compression cycles.

To quantify the metastability, we computed a measure $\xi = \langle (h(x,n+2) - h(x,n))^2 \rangle_x$, where $\langle .. \rangle_x$ corresponds to 
average over $x$, and identify the rearrangements by finding the number of peaks in $\xi$ above a given threshold. The number $\langle N_\xi \rangle$ in the first, second and third part of the 1000 cycles is plotted in Fig.~\ref{fig:sec_evo}(d) averaged over three trials in the case of each $\theta_f$. It shows that the rearrangements for $\theta_f \leq 540^\circ$ eventually decrease and the folded shapes converge over time even as the creases sharpen. However, for large enough deformations corresponding to $\theta_f = 720^\circ$, the rearrangements can continue to occur even as they appear to converge over hundreds of cycles. In fact, we find that rearrangements persist in the case of $\theta_f = 720^\circ$ even as
the sheet tears at the creases due to the weakening caused by repeated plastic deformations after about $n=1800$ (see Supplementary Document~\cite{sup-doc}). Thus, a recurrent folded structure does not always exist for large enough deformations as fatigue leads the sheet to progressively tear up and its genus changes.  

The observed overall progress toward a limit cycle in crumpled sheets over a transient number of cycles that increase with increased $\theta_f$ is similar to that observed in cyclically sheared amorphous materials~\cite{Regev2013}. There, a relaxation to a limit cycle, and transitions between limit cycles was also observed with applied shear amplitude. For large enough amplitudes only irreversible migration of constituent particles was observed~\cite{Regev2013,Pine2005}. In particular the convergence of the folded structure for sufficiently low $\theta_f$, but irreversibility for high enough $\theta_f$, can be noted to be similar to those observed in multi-body systems. Our system, with the slow evolution of the curvature of the folds with training cycle, further indicates that the broad features of reversibility and irreversibility with applied amplitude can survive sufficiently slow evolution of fatigue in the system.     

\section{Conclusions}

In summary, we find that a repeatedly crumpled sheet can perform reversible transformations even as they are crushed significantly over a wide range of applied loading amplitudes. 
Even when the system displays metastability and continuous logarithmic evolution, we find that a convergent structure can be reached after a sufficiently large number of training cycles depending on the depth of the compression cycle. However, for sufficiently large loading and compaction, a convergent structure is not reached as the sheet fatigues and its topology changes as it tears. Overall, we find that the behavior of crumpled sheet under repeated loading is similar to other disordered system, and in particular the reversible transformation of athermal particles subjected to cyclic shear~\cite{Pine2005,Regev2013,Paulsen2014}. There, reversibility is observed for a range of sufficiently small strain amplitude, but irreversibility develops with increasing amplitude depending on the Lyapunov exponents of the system. While, a similar analysis is beyond the scope of this study, it may be anticipated that our results can stimulate work in that direction.

%\begin{center}
%{\bf Acknowledgements}
%\end{center}

\begin{acknowledgments}
We thank Sohum Kapadia, Craig Maloney, Alex Petroff, and Alessio Zaccone for constructive comments on the manuscript, and Jovana Andrejevic and Animesh Biswas for discussions while the study was being developed.  This work was supported under U.S. National Science Foundation grant DMR-2005090. 
\end{acknowledgments}

\appendix
 
\section{Spatial distribution of crumples}
\label{sec:curv}
\begin{figure}[h]
 \includegraphics[width=8cm]{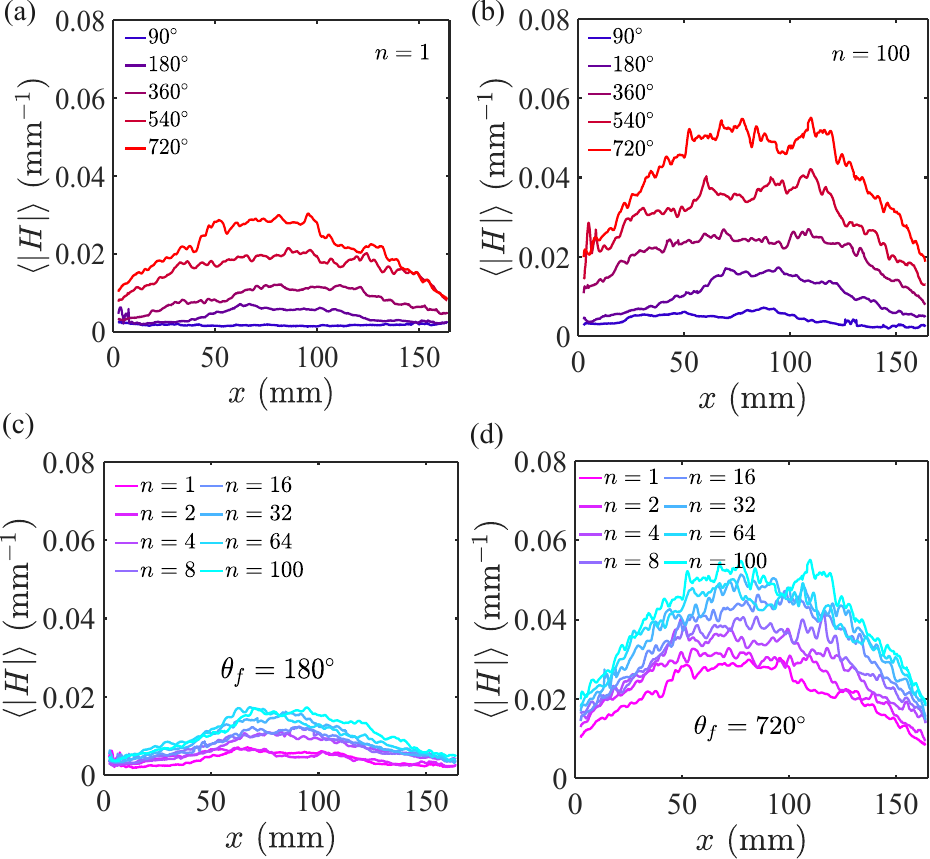}
    \caption{ Mean curvature of the crumpled sheet as a function of distance along the sheet's length for various $\theta_{f}$ after cycle $n=1$ (a)  and $n=100$ (b), respectively. (c,d) Mean curvature of the cross section as a function of distance along the sheet's length for increasing $n$ for $\theta_{f} = 180^{\circ}$ and $720^{\circ}$, respectively. The sheet is more crumpled in the central regions between the clamps where sheet forms a tight bundle when twisted. }
    \label{fig:Curv_throughout_length}
\end{figure}

We plot the average $|H|$ as a function of distance from one clamped end to the other in Fig.~\ref{fig:Curv_throughout_length}. 
As can be seen from Fig.~\ref{fig:Curv_throughout_length}(a,b), $H$ %averaged over constant distance from the clamps 
is lower near in the clamped edges because the sheet is forced to maintain the curvature of the end disks on which it is mounted. %, which is $1/26.4$\,mm$^{-1}$. 
When the sheet is repeatedly twisted, $H$ increase systematically everywhere and thus the variation across $x$ persists (see Fig.~\ref{fig:Curv_throughout_length}(c,d)). Accordingly, we analyze the distribution of the measured $H$ over a $146$\,mm by $115$\,mm centrally cropped area of the sheet to avoid direct boundary effects.

%\newpage 

\section{Ridge angle distribution}
\label{sec:angle}
\begin{figure}[h]
    \centering
             \includegraphics[width=8cm]{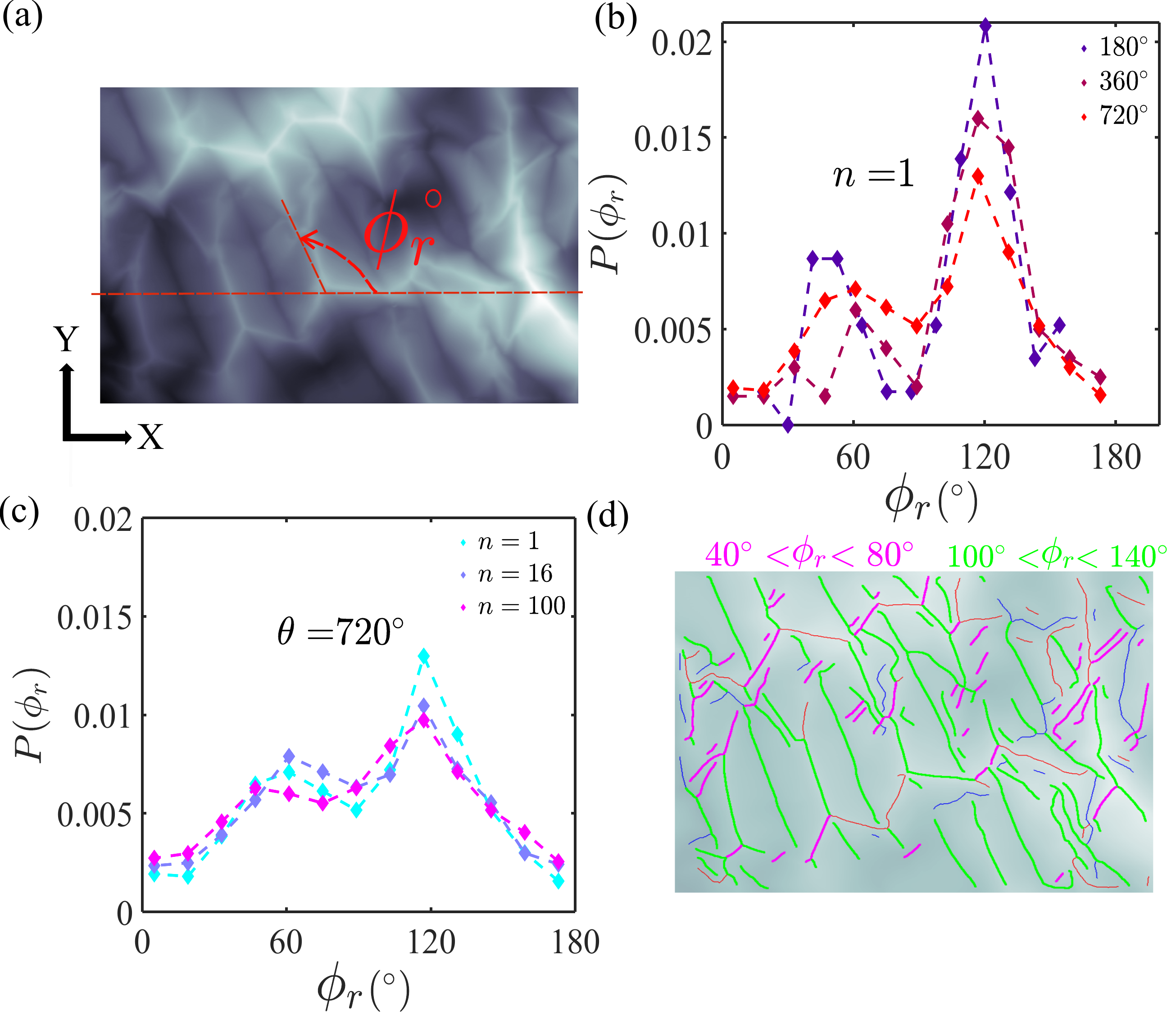}
    \caption{ (a) A schematic of the orientation angle of ridge $\phi_r$ w.r.t. $x$-axis. The distribution of ridge angles w.r.t. the $x$-axis is observed to anisotropic. The distributions become somewhat broader with increasing twist angle $\theta$ (b), and loading cycles $n$ (c). (d) The creases contributing to the peaks in the ridge angle distribution graph are highlighted with magenta and green colors. Magenta for angles between $40^\circ$ and $80^\circ$, and green for angles between $100^\circ$ and $140^\circ$. }
    \label{fig:ridgeangle}
\end{figure}
We obtained the orientation angle the ridges formed relative to the $x$-axis which is orthogonal the axis of rotation. Figure~\ref{fig:ridgeangle} shows the distributions observed with increasing $\theta$ and $n$. We find that the distributions are not flat, but rather has two broad peaks related to the wrinkles which are formed following the primary buckling of the sheet. The broad peak at $\phi_r \approx 120^\circ$ corresponds the initial wrinkles which form after the primary buckling instability. These wrinkles then crease into ridges and valleys which are denoted by green lines. The second peak at $\phi_r \approx 60^\circ$ corresponds to shorter creases which form between the primary creases with continuing twist and are denoted in magenta.  The peaks are observed to broaden further with increasing $n$ and with increasing applied twist. 

%\bibliographystyle{unsrt}
%\bibliography{crumple}

\begin{thebibliography}{10}

\bibitem{Amar1997}
M.~Ben Amar and Y.~Pomeau.
\newblock Crumpled paper.
\newblock {\em Proceedings: Mathematical, Physical and Engineering Sciences},
  453(1959):729--755, 1997.

\bibitem{Blair2005}
Daniel~L. Blair and Arshad Kudrolli.
\newblock Geometry of crumpled paper.
\newblock {\em Phys. Rev. Lett.}, 94:166107, Apr 2005.

\bibitem{Boue2006}
L.~Bou\'e, M.~Adda-Bedia, A.~Boudaoud, D.~Cassani, Y.~Couder, A.~Eddi, and
  M.~Trejo.
\newblock Spiral patterns in the packing of flexible structures.
\newblock {\em Phys. Rev. Lett.}, 97:166104, Oct 2006.

\bibitem{Witten2007}
T.~A. Witten.
\newblock Stress focusing in elastic sheets.
\newblock {\em Rev. Mod. Phys.}, 79:643--675, Apr 2007.

\bibitem{Croll2019}
Andrew Croll, Timothy Twohig, and Theresa Elder.
\newblock The compressive strength of crumpled matter.
\newblock {\em Nature Communications}, 10:1502, 04 2019.

\bibitem{Matan2002}
Kittiwit Matan, Rachel~B. Williams, Thomas~A. Witten, and Sidney~R. Nagel.
\newblock Crumpling a thin sheet.
\newblock {\em Phys. Rev. Lett.}, 88:076101, Jan 2002.

\bibitem{Fokker2019}
M.~C. Fokker, S.~Janbaz, and A.~A. Zadpoor.
\newblock Crumpling of thin sheets as a basis for creating mechanical
  metamaterials.
\newblock {\em RSC Adv.}, 9:5174--5188, 2019.

\bibitem{Venkataramani2019}
Shankar~C. Venkataramani.
\newblock Buckling sheets open a door to understanding self-organization in
  soft matter.
\newblock {\em Proceedings of the National Academy of Sciences},
  116(5):1477--1479, 2019.

\bibitem{Persson1998}
B.N.J. Persson.
\newblock {\em Sliding Friction: Physical Principles and Applications}.
\newblock NanoScience and Technology. Springer Berlin Heidelberg, 2013.

\bibitem{Bertin2003}
E.~Bertin, Jean-Philippe Bouchaud, J-M Drouffe, and C~Godreche.
\newblock The kovacs effect in model glasses.
\newblock {\em Journal of Physics A General Physics}, 36, 06 2003.

\bibitem{Thiria2011}
B.~Thiria and M.~Adda-Bedia.
\newblock Relaxation mechanisms in the unfolding of thin sheets.
\newblock {\em Phys. Rev. Lett.}, 107:025506, Jul 2011.

\bibitem{Amir2012}
Ariel Amir, Yuval Oreg, and Yoseph Imry.
\newblock On relaxations and aging of various glasses.
\newblock {\em Proceedings of the National Academy of Sciences},
  109(6):1850--1855, 2012.

\bibitem{Gottesman2018}
Omer {Gottesman}, Jovana {Andrejevic}, Chris~H. {Rycroft}, and Shmuel~M.
  {Rubinstein}.
\newblock {A state variable for crumpled thin sheets}.
\newblock {\em Communications Physics}, 1(1):70, 2018.

\bibitem{Keim2019}
Nathan~C. Keim, Joseph~D. Paulsen, Zorana Zeravcic, Srikanth Sastry, and
  Sidney~R. Nagel.
\newblock Memory formation in matter.
\newblock {\em Rev. Mod. Phys.}, 91:035002, Jul 2019.

\bibitem{DiDonna2001}
B.~A. DiDonna and T.~A. Witten.
\newblock Anomalous strength of membranes with elastic ridges.
\newblock {\em Phys. Rev. Lett.}, 87:206105, Oct 2001.

\bibitem{Santangelo2017}
Christian~D. Santangelo.
\newblock Extreme mechanics: Self-folding origami.
\newblock {\em Annual Review of Condensed Matter Physics}, 8(1):165--183, 2017.

\bibitem{Ribes2024}
Gerard Giménez-Ribes, Aref Ghorbani, Soon~Yuan Teng, Erik {van der Linden},
  and Mehdi Habibi.
\newblock Shear and shear-induced normal responses of origami cylinders relate
  to their structural asymmetries.
\newblock {\em Materials \& Design}, 240:112874, 2024.

\bibitem{Stern2020}
Menachem Stern, Chukwunonso Arinze, Leron Perez, Stephanie~E. Palmer, and
  Arvind Murugan.
\newblock Supervised learning through physical changes in a mechanical system.
\newblock {\em Proceedings of the National Academy of Sciences},
  117(26):14843--14850, 2020.

\bibitem{Stern2018}
Menachem Stern, Viraaj Jayaram, and Arvind Murugan.
\newblock Shaping the topology of folding pathways in mechanical systems.
\newblock {\em Nature Communications}, 9:4303, 2018.

\bibitem{Arinze2023}
Chukwunonso Arinze, Menachem Stern, Sidney~R. Nagel, and Arvind Murugan.
\newblock Learning to self-fold at a bifurcation.
\newblock {\em Phys. Rev. E}, 107:025001, Feb 2023.

\bibitem{Pine2005}
David Pine, J~Gollub, J~Brady, and Alexander Leshansky.
\newblock Chaos and threshold for irreversibility in sheared suspensions.
\newblock {\em Nature}, 438:997--1000, 2005.

\bibitem{Regev2013}
Ido Regev, Turab Lookman, and Charles Reichhardt.
\newblock Onset of irreversibility and chaos in amorphous solids under periodic
  shear.
\newblock {\em Phys. Rev. E}, 88:062401, Dec 2013.

\bibitem{Shohat_2022}
Dor Shohat, Daniel Hexner, and Yoav Lahini.
\newblock Memory from coupled instabilities in unfolded crumpled sheets.
\newblock {\em Proceedings of the National Academy of Sciences}, 119(28), 2022.

\bibitem{Jules2022}
Th\'eo Jules, Austin Reid, Karen~E. Daniels, Muhittin Mungan, and Fr\'ed\'eric
  Lechenault.
\newblock Delicate memory structure of origami switches.
\newblock {\em Phys. Rev. Res.}, 4:013128, Feb 2022.

\bibitem{Xia2022}
Yutong Xia, Narayanan Kidambi, Evgueni Filipov, and K.~W. Wang.
\newblock {Deployment Dynamics of Miura Origami Sheets}.
\newblock {\em Journal of Computational and Nonlinear Dynamics}, 17(7):071005,
  04 2022.

\bibitem{Dunne2022}
Francis Dunne, Kjell Westra, Caleb Mattox, and Jacob Leachman.
\newblock Fatigue life characterization of hand folded and vacuum formed
  kresling origami bellows at 77k.
\newblock {\em IOP Conference Series: Materials Science and Engineering},
  1241(1):012014, may 2022.

\bibitem{Tallinen2009}
T.~Tallinen, J.~A. {\AA}str{\"o}m, and J.~Timonen.
\newblock The effect of plasticity in crumpling of thin sheets.
\newblock {\em Nature materials}, 8 1:25--9, 2009.

\bibitem{Habibi2017}
Mehdi Habibi, Mokhtar Adda-Bedia, and Daniel Bonn.
\newblock Effect of the material properties on the crumpling of a thin sheet.
\newblock {\em Soft Matter}, 13:4029--4034, 2017.

\bibitem{Rahmayanti2016}
Handika~Dany Rahmayanti, Fisca~Dian Utami, and Mikrajuddin Abdullah.
\newblock Physics model for wringing of wet cloth.
\newblock {\em European Journal of Physics}, 37(6):065806, sep 2016.

\bibitem{Dong2023}
Pan Dong, Mengfei He, Nathan~C. Keim, and Joseph~D. Paulsen.
\newblock Twisting a cylindrical sheet makes it a tunable locking material.
\newblock {\em Phys. Rev. Lett.}, 131:148201, Oct 2023.

\bibitem{Houle1996}
Paul~A. Houle and James~P. Sethna.
\newblock Acoustic emission from crumpling paper.
\newblock {\em Phys. Rev. E}, 54:278--283, Jul 1996.

\bibitem{sup-doc}
See supplemental material at [url will be inserted by publisher] for further
  information on methods, movies and analysis.

\bibitem{Donnell1935}
L.~H. Donnell.
\newblock Stability of thin-walled tubes under torsion.
\newblock Technical report, National Advisory Committee for Aeronautics -
  Report No. 479, 1933.

\bibitem{HUNT2005}
Giles~W. Hunt and Ichiro Ario.
\newblock Twist buckling and the foldable cylinder: an exercise in origami.
\newblock {\em International Journal of Non-Linear Mechanics}, 40(6):833--843,
  2005.

\bibitem{Wang2020}
Li-Min Wang, Sun-Ting Tsai, Chih-yu Lee, Pai-Yi Hsiao, Jia-Wei Deng, Hung-Chieh
  Fan~Chiang, Yicheng Fei, and Tzay-Ming Hong.
\newblock Crumpling-origami transition for twisting cylindrical shells.
\newblock {\em Phys. Rev. E}, 101:053001, May 2020.

\bibitem{Ribes2023}
Gerard Giménez-Ribes, Melika Motaghian, Erik {van der Linden}, and Mehdi
  Habibi.
\newblock Crumpled structures as robust disordered mechanical metamaterials.
\newblock {\em Materials \& Design}, 232:112159, 2023.

\bibitem{Chopin2022}
Julien Chopin and Arshad Kudrolli.
\newblock Tensional twist-folding of sheets into multilayered scrolled yarns.
\newblock {\em Science Advances}, 8(14):eabi8818, 2022.

\bibitem{Jules2020}
T.~Jules, F.~Lechenault, and M.~Adda-Bedia.
\newblock Plasticity and aging of folded elastic sheets.
\newblock {\em Phys. Rev. E}, 102:033005, Sep 2020.

\bibitem{Shohat_2023}
Dor Shohat, Yaniv Friedman, and Yoav Lahini.
\newblock Logarithmic aging via instability cascades in disordered systems.
\newblock {\em Nature Physics}, 19(12):1890–1895, 2023.

\bibitem{Paulsen2014}
Joseph~D. Paulsen, Nathan~C. Keim, and Sidney~R. Nagel.
\newblock Multiple transient memories in experiments on sheared non-brownian
  suspensions.
\newblock {\em Phys. Rev. Lett.}, 113:068301, Aug 2014.

\bibitem{Cambou2011}
Anne~Dominique Cambou and Narayanan Menon.
\newblock Three-dimensional structure of a sheet crumpled into a ball.
\newblock {\em Proceedings of the National Academy of Sciences},
  108(36):14741--14745, 2011.

\bibitem{Deboeuf2013}
S.~Deboeuf, E.~Katzav, A.~Boudaoud, D.~Bonn, and M.~Adda-Bedia.
\newblock Comparative study of crumpling and folding of thin sheets.
\newblock {\em Phys. Rev. Lett.}, 110:104301, Mar 2013.

\bibitem{Boudaoud2000}
Arezki Boudaoud, Pedro Patrício, Yves Couder, and Martine Ben~Amar.
\newblock Dynamics of singularities in a constrained elastic plate.
\newblock {\em Nature}, 407:718--20, 11 2000.

\end{thebibliography}

\end{document}